# Light-induced instabilities in photo-oriented liquid crystal cells

I. Jánossy, K. Fodor-Csorba, A. Vajda, L.O. Palomares and T. Tóth-Katona

*Institute for Solid State Physics and Optics, Wigner Research Centre for Physics,*

*Hungarian Academy of Sciences, H-1525 Budapest, P.O.Box 49, Hungary*

e-mails

I. Janossy:        janossy.istvan@wigner.mta.hu

K. Fodor-Csorba:   fodor.tamasne@wigner.mta.hu

A. Vajda:          vajda@szfki.hu

L.O. Palomares:    lpaloma_quark@yahoo.com.mx

T. Tóth-Katona:    tothkatona.tibor@wigner.mta.hu

Running title: Light-induced instabilities…

Abstract

In a planar liquid crystal sample sandwiched between a photosensitive and a reference plate instabilities occurred, when the cell was illuminated from the reference side. The instabilities were induced both by polarized white light source and monochromatic laser beams. Static and dynamic regimes were found; for laser irradiation dynamic instability was found only in a range of polarization directions. A model, developed for monochromatic excitation, predicts that at certain thicknesses dynamic instability is forbidden. Experiments on a wedge-like cell confirmed this conclusion.

1. Introduction

Photoalignment of nematic liquid crystals has been studied throughout the last two decades [1,2,3]. The process is interesting both from the points of view of basic research and of technical applications. In a typical experiment a sandwich-cell is prepared from the liquid crystal. One of the substrates is made photosensitive that allows controlling the director alignment on it by an external light source, for instance via the polarization direction. The other substrate is a traditionally prepared plate that ensures fixed director orientation on it (reference plate). On illuminating the photosensitive plate the director pattern can be changed in the cell, e.g. from planar orientation to a twisted one.

In current publications [4,5] we showed that when a planar cell is illuminated from the reference face instabilities occur. The first observations were made in a polarizing microscope, where the exciting light source was the white lamp of the microscope itself. In the largest part of the samples a static pattern was detected; however, at certain locations of the cell a dynamic instability occurred. In a very recent paper the investigations were extended for the case of monochromatic irradiation from a polarized laser beam. Here also static and dynamic instabilities were observed. The character of the pattern in this case depended on the direction of light polarization rather then on the location of the illuminated area. In Section 2 we summarize the main features of the instabilities observed for the two set-ups.

In Ref. [5] a model was proposed for the laser-induced instabilities, which we review in Section 3. The model predicts that when in a planar cell the phase shift between the ordinary and extraordinary beams of the exciting light is $(m+\frac{1}{2})\pi$ ($m$ integer) the instability should be static at any direction of the polarization. We present observations on wedge-like cells, in which the above condition was fulfilled along particular lines in the cell. In agreement with the model, when the whole sample was irradiated by a laser light we did not observe dynamic scattering in the vicinity of these lines. The details are presented in Section 4.

2. Experimental observations

The fabrication of the azo-dye monolayer coated photosensitive plate was described earlier [4,6]. As reference plates rubbed polyimide coated slides from E.H.C. Co (Japan) were used. 10-15 µm thick cells have been filled with the room temperature nematic 4-cyano-4'-pentylbiphenyl (5CB). Before and during the filling process the cell was illuminated from the photosensitive side with light from a white LED source, polarized perpendicularly to the rubbing direction. The procedure resulted in good quality planar alignment.

The initial observations were carried out in a polarizing microscope, where the illuminating beam was the white-light source of the microscope [4]. When the dye-coated plate was facing the illuminating beam (direct geometry), a normal reorientation process was observed. The reorientation from planar to twisted or twisted to planar configuration was induced by rotating the polarizer of the microscope. It took place in several seconds at low illumination levels (0.05 mW/mm$^2$) and in less than a second at high illumination levels (0.5 mW/mm$^2$). By rotating the polarizer over 90 degrees a super-twist as high as 360 degrees could be generated in the sample; however, at such high super-twists, after a few seconds disclination loops were formed at different points of the sample. These loops propagated through the illuminated area, restoring the planar configuration.

When the reference plate was facing the illuminating beam (reverse geometry) instabilities of the planar configuration were observed. Two kinds of instabilities were detected. In the major part of the sample a static pattern was formed, with a fine structure on the scale of a micrometer (Figure 1). At some particular locations of the samples, mainly in the vicinity of the cell edge, dynamical instability occurred (Figure 2). Within these regions undamped and chaotic director rotation took place. The details are described in [4].

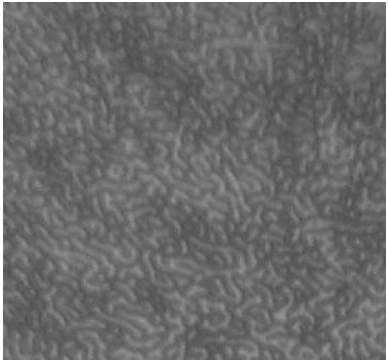 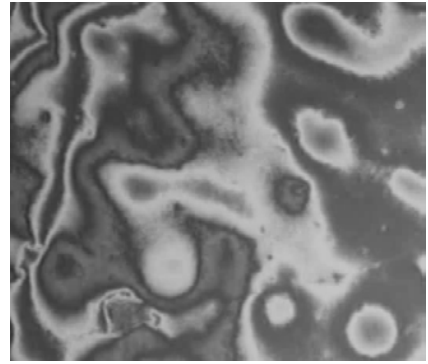

Figure 1. Static pattern observed in a polarizing microscope

Figure 2. Dynamic instability observed in a polarizing microscope

The case of a polarized monochromatic laser beam as the exciting light source was studied recently [5]. Two laser sources were used with the wavelength 532 nm (green) and 457 nm (blue). In the direct geometry standard reorientation process was observed, similar to the one detected in a polarizing microscope. For the reverse geometry again instabilities were found. The character of the instability, however, in this case depended on the angle $\alpha$ between the light polarization and the director orientation on the reference cell. The following cases were found ($\alpha = 0$ corresponds to the light polarization perpendicular to the rubbing direction).

1. $\alpha = 0$. The planar configuration remained stable, no depolarization or light scattering took place.
2. $\alpha = 90°$. The light beam became partially depolarized and a faint scattered halo appeared behind the sample, with an aperture of about 20-30 degrees for the green laser. The scattered light was steady in time and obviously corresponded to the static domain structure, observed in the microscope.

The domain size estimated from the aperture was about a few micrometers, which agrees with the pattern period induced with the microscope light.

3. $0 < \alpha_1 < \alpha < \alpha_2 < 90°$. In this geometry dynamic light scattering occurred. It was difficult to determine the precise values of the angles $\alpha_1$ and $\alpha_2$; a typical value is 15° and 75° respectively. This dynamic scattering apparently is related to the dynamic instability seen in the microscope. In the case of laser excitation, however, it was observed in the main part of the samples, in contrast with the case of white light irradiation, when the turbulent behaviour developed only towards the boundaries of the cell.

The light scattering for cases 2 and 3 are illustrated in Figure 3. A detailed analysis of the fluctuations in the scattered light is given in [5].

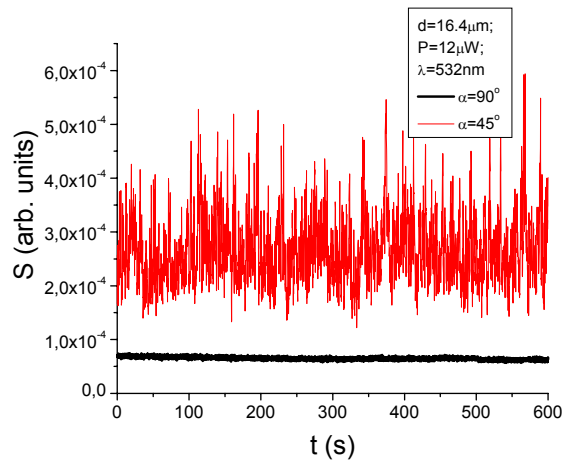

Figure 3. Temporal dependence of the intensity fluctuations of the scattered light (S) for the polarizations α=90° and α=45° of the incoming light in the reverse geometry.

3. Model for the laser-induced instability

The process of photoorientaton is connected with the *trans-cis* isomerization of the azo molecules attached to the light-sensitive plate. Initially the dye molecules are in *trans* configuration. For linearly polarized light, orientationally selective excitation converts the dyes aligned parallel to the light polarization to *cis* isomers. Therefore the orientational distribution of the *trans* molecules becomes anisotropic, with a peak along the direction perpendicular to the polarization. This direction becomes also the "easy axis" of the surface alignment of the director on the azo-dye coated plate. For elliptically polarized light the peak of the distribution is along the minor axis of the ellipse, therefore this direction becomes the easy axis.

In the following we present a simplified account on the origin of the instabilities for the case of laser excitation. The exact description of the dynamics of the photoorientation and the associated reorientation of the nematic liquid crystal layer in the cell is a complex task that has not yet been solved completely. Here we use a simplified approach, which is based on surface torques acting on the photosensitive plate. We assume that a light-induced torque acts on the surface director, which has the same functional form as the bulk light-induced torque:

$$\Gamma_{ph} = -f <(\mathbf{E}_s \times \mathbf{n}_s)\mathbf{E}_s \cdot \mathbf{n}_s > \quad /1/$$

where $\mathbf{E}_s$ and $\mathbf{n}_s$ are the surface electric field and the director, respectively; $f$ is a phenomenological parameter and $<>$ denotes averaging for periods of oscillation of the electromagnetic wave. This torque is opposed by the elastic torque acting on the surface director. As in our experiments both the director and the electric field is in the plane of the substrates, the two surface torques are normal to the plate. The above expression is valid for elliptically polarized light too.

In the following we analyze different cases for the polarization angle, $\alpha$. We assume that the initial director configuration is planar. When the input light on the reference plate is linearly polarized at an angle $\alpha$, in the initial planar cell the photoinduced surface torque becomes

$$\Gamma_{ph} = -f'I\sin 2\alpha \cos\Delta\Phi \quad /2/$$

where I is the light intensity, $\Delta\Phi = 2\pi(n_e - n_o)d/\lambda$ is the phase difference between the extraordinary and ordinary components of the wave at the photosensitive plate; $n_e$ and $n_o$ are the extraordinary and ordinary refractive index, respectively; $d$ and $\lambda$ are the sample thickness and wavelength respectively; $f'$ is a constant related to $f$.

a. $\alpha = 0$. In this case the photoinduced torque is zero and the surface director on the dyed plate is stable against small fluctuations. As a result the planar configuration remains stable on illumination.

b. $0 < \alpha < 90°$, $\cos\Delta\Phi \neq 0$. In this case the initial photoinduced torque is different from zero; therefore the surface director rotates away from its original orientation. On the other side, the light polarization ellipse rotates as well. In the so-called Mauguin limit the magnitude of the principal axes of the ellipse remains the same, but they rotate together with the director [7,8]. In this limit the photoinduced surface torque does not change during the director rotation. The elastic torque is increasing as the director configuration becomes twisted. However, as shown by experiments in the direct geometry, the latter torque does not balance the former one even at twist angles where an inversion wall is formed in the cell. The inversion wall reduces the twist angle, therefore the elastic torque diminishes. Therefore the director rotation continues, never arriving at a static state. This mechanism explains the observed dynamic scattering. The process is sketched in Figure 4.

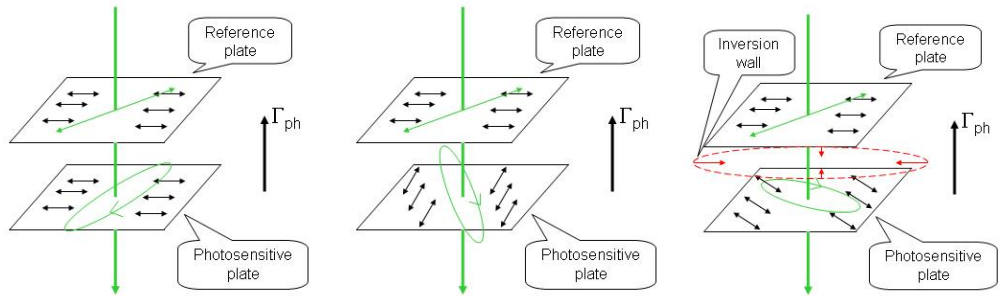

Figure 4. Rotation of the director and the polarization ellipse on the photosensitive plate from the initial planar configuration until the formation of an inversion wall.

c. $\alpha = 90°$. The photoinduced torque is zero, but the surface director is unstable against small fluctuations. The fluctuations in the present case imply static orientational disorder on the reference plate rather than thermal fluctuations of the director field [9]. When the sample is illuminated, the local surface director on the dyed plate rotates clockwise or anti-clockwise, depending on the sign of the deviation of the orientation from the average one at the particular location. We suggest that the formation of domains reduces significantly the photoinduced torque, meanwhile increases the elastic one. As a result static configuration can develop in this geometry.

d. $0 < \alpha < 90°$, $\cos \Delta\Phi = 0$. This situation takes place when the thickness of the layer fulfills the condition

$$d = (m + \frac{1}{2})\lambda / 2(n_e - n_o) \quad /3/$$

where $m$ is an integer. In this case the photoinduced torque is zero. The director on the photosensitive plate is stable against small fluctuations, so dynamic scattering is not expected to occur at such thicknesses. In the next Section we provide experimental evidence of this conclusion.

4. <u>Investigation of a wedge-like cell</u>

In order to study the thickness dependence of the laser-induced instability, we prepared a cell with varying thickness. The thickness was scanned with the help of a spectrophotometer across the cell before filling in the liquid crystal. The thickness decreased from one spacer to the other one from 30 µm to 22 µm at the top of the sample and from 25 µm to 23 µm at the bottom of it. $\Delta\Phi$ was determined using the known data for the refractive indices of 5CB [10]. For the wavelength 457 nm the locations at which the condition /3/ is fulfilled where also directly measured in the filled cell with the help of a photo-elastic modulator (PEM); the second harmonic signal from the PEM disappears at these points. The coincidence between the locations determined by the two methods was satisfactory.

The entire cell was irradiated by a laser beam from the reference side at $\alpha = 20°$. Dynamic scattering caused the sample to become opaque due to the disclination loops generated in it by the light-induced instability. In Figure 4 a photograph is shown of the sample after irradiating it with 532nm. As can be seen from the figure, stripes were formed, in which no disclination loops were present. The filled circles shown in the figure indicate the locations where the condition cos $\Delta\Phi = 0$ is satisfied for the wavelength of illumination. These are within, or very near to the observed stripes, supporting the validity of our model.

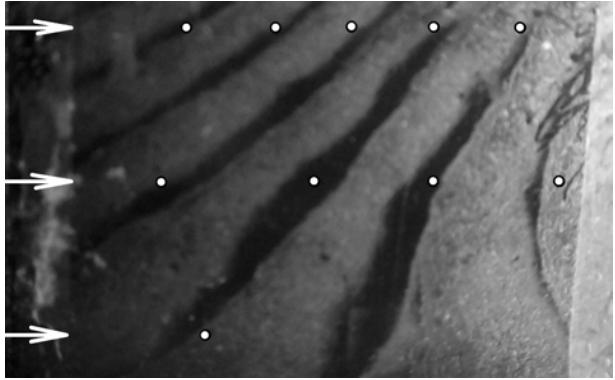

Figure 5. A wedge-like sample after irradiation with 532nm. The white spots show the locations for three different rows were the condition /3/ is fulfilled for the irradiating laser wavelength 532nm.

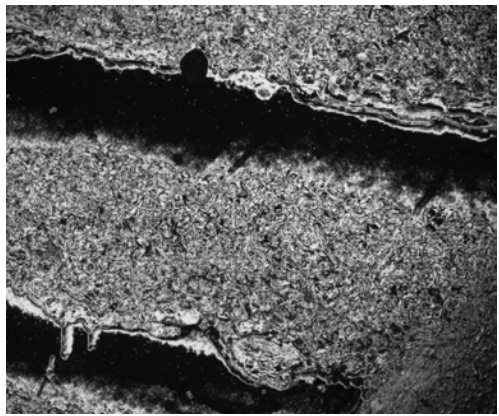

Figure 6. Part of the sample in a polarizing microscope after irradiation at 532 nm.

From Figure 5 it is apparent that the stripes exhibit sharp edges. This circumstance is reinforced in Figure 6, where a photograph taken in a polarizing microscope is shown.

The sharp boundaries indicate that dynamic scattering is forbidden at a certain range around the thicknesses obeying condition /3/. When we irradiated the sample with a wavelength 457 nm, the predicted points were again situated close to the stripes (Figure 7), although the stripes became much narrower than in the case of the green laser. These observations suggest that there is a threshold value of $\Delta\Phi$ for the occurrence of dynamic scattering. The threshold may depend on $\alpha$, the light intensity, the wavelength and on the extent of disorder of orientation on the reference plate. Another way to interpret this fact is that the angles $\alpha_1$ and $\alpha_2$, introduced in Section 2, depend on $\Delta\Phi$ and the other factors mentioned before. This point needs further investigations.

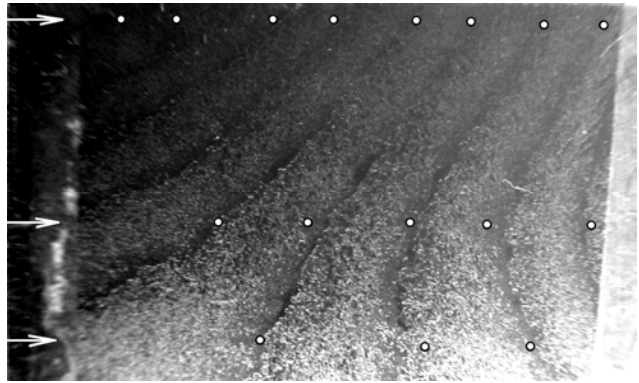

Figure 7. The sample after irradiation with 457nm. The white spots show the locations for three different rows were the condition /3/ is fulfilled for the irradiating laser wavelength 457nm.

5. <u>Summary</u>

In the paper we described ligth-induced instabilities in nematic liquid crystal cells. The reason of pattern formation in the reverse geometry is the coupling between the photoalignment and director reorientation. The reorientation influences the light polarization state at the photosensitive plate and thus the photoorientation; in turn, the director reorientation depends on the photo-induced alignment. This interaction leads to the observed instabilities.

The effect described in the paper can be used to create opaque spots in an otherwise transparent sample at well localized areas. These spots can be erased by thermal treatment of the sample, thus yielding the possibility for a rewriteable optical storage device.

Acknowledgment

The work was supported by the Hungarian Research Fund OTKA K81250.

References

[1] W.M. Gibbons, P.J. Shannon, S.T. Sun, B.J. Swetlin (1991) *Nature 351*, 49.

[2] K. Ichimura (2000) *Chem. Rev. 100,* 1847

[3] V. G. Chigrinov, V. M. Kozenkov, Hoi-Sing Kwok (2008) *Photoalignment of Liquid Crystalline Materials: Physics and Applications* John Wiley & Sons.

[4] I. Jánossy, K. Fodor-Csorba, A. Vajda, L. Palomares (2011) *Appl. Phys. Lett. 99*, 111103.

[5] I. Jánossy, K. Fodor-Csorba, A. Vajda, T. Tóth-Katona (2013) submitted to *Phys. Rev. E.*

[6] Y. Yi, M. J. Farrow, E. Korblova, D. M. Walba, and E. Furtak (2009) *Langmuir 25*, 997.

[7] M.C. Mauguin (1911) *Bull. Soc. Fr. Mineral. 34*, 71.

[8] H. de Vries (1951) *Acta Cryst. 4*, 219.

[9] M. Nespoulous, Ch. Blanc, M. Nobili (2010) *Phys. Rev. Lett. 104*, 097801.

[10] S.T. Wu, R.J. Cox (1988) *Journ. Appl. Phys. 64*, 821.